\documentstyle[12pt]{article}
\input epsf
\epsfverbosetrue
\newlength{\dinwidth} 
\newlength{\dinmargin}
\begin{document}

DESY 96-094

May 1996

\def\cts{\cos\theta^{\ast}}
\def\xgo{x_\gamma^{OBS}}
\def\xpo{x_p^{OBS}}
\def\xg{x_\gamma}
\def\ETJ{E_T^{jet}}
\def\ETAJ{\eta^{jet}}
\def\ETAB{\bar{\eta}}
\def\EEP{E^\prime_{e}}
\def\TEP{\theta^\prime_{e}}
\def\MJJ{M_{jj}}

\thispagestyle{empty}

\begin{center}
{\Huge Dijet Angular Distributions in Direct and Resolved Photoproduction at HERA}
\end{center}

\vspace{1cm}

\begin{center}
\begin{Large}
ZEUS Collaboration\\
\vspace{1cm}
\end{Large}
\end{center}

\setcounter{page}{0}

\vspace{4cm}

\begin{abstract}

Jet photoproduction, where the two highest transverse energy ($\ETJ$) jets 
have $\ETJ$ above 6 GeV and a jet-jet invariant mass above 23~GeV, has 
been studied with the ZEUS detector at the HERA $ep$ collider.
Resolved and direct photoproduction samples have been separated.
The cross section as a function of the angle between the jet-jet 
axis and the beam direction in the dijet rest frame has been 
measured for the two samples. 
The measured angular distributions differ markedly from each other. 
They agree with the predictions of QCD calculations, where the different
angular distributions reflect the different spins of the quark and gluon 
exchanged in the hard subprocess.
\end{abstract} 

\textwidth17.0cm

\newpage

\topmargin-1.cm                                                                                    
\evensidemargin-0.3cm                                                                              
\oddsidemargin-0.3cm                                                                               
\textwidth 16.cm                                                                                   
\textheight 680pt                                                                                  
\parindent0.cm                                                                                     
\parskip0.3cm plus0.05cm minus0.05cm

\def\3{\ss}                                                                                        
\newcommand{\address}{ }                                                                           
\renewcommand{\author}{ }                                                                          
\pagenumbering{Roman}
                                                   %
\begin{center}                                                                                     
{                      \Large  The ZEUS Collaboration              }                               
\end{center}                                                                                       

  M.~Derrick,                                                                                      
  D.~Krakauer,                                                                                     
  S.~Magill,                                                                                       
  D.~Mikunas,                                                                                      
  B.~Musgrave,                                                                                     
  J.R.~Okrasinski,                                                                                 
  J.~Repond,                                                                                       
  R.~Stanek,                                                                                       
  R.L.~Talaga,                                                                                     
  H.~Zhang  \\                                                                                     
 {\it Argonne National Laboratory, Argonne, IL, USA}~$^{p}$
% ...
                                                           %
                                                           %
% ...
$^{  20}$ now at ATLAS Collaboration, Univ. of Munich \\                                           
% ...
                                                           %
                                                           %
                                                           %
                                                           %
\begin{tabular}[h]{rp{14cm}}                                                                       
$^{a}$ &  supported by the Natural Sciences and Engineering Research                               
          Council of Canada (NSERC)  \\                                                            
$^{b}$ &  supported by the FCAR of Qu\'ebec, Canada  \\                                            
$^{c}$ &  supported by the German Federal Ministry for Education and                               
          Science, Research and Technology (BMBF), under contract                                  
          numbers 056BN19I, 056FR19P, 056HH19I, 056HH29I, 056SI79I \\                              
$^{d}$ &  supported by the MINERVA Gesellschaft f\"ur Forschung GmbH,                              
          the Israel Academy of Science and the U.S.-Israel Binational                             
          Science Foundation \\                                                                    
$^{e}$ &  supported by the German Israeli Foundation, and                                          
          by the Israel Academy of Science  \\                                                     
$^{f}$ &  supported by the Italian National Institute for Nuclear Physics                          
          (INFN) \\                                                                                
$^{g}$ &  supported by the Japanese Ministry of Education, Science and                             
          Culture (the Monbusho) and its grants for Scientific Research \\                         
$^{h}$ &  supported by the Korean Ministry of Education and Korea Science                          
          and Engineering Foundation  \\                                                           
$^{i}$ &  supported by the Netherlands Foundation for Research on                                  
          Matter (FOM) \\                                                                          
$^{j}$ &  supported by the Polish State Committee for Scientific                                   
          Research, grants No.~115/E-343/SPUB/P03/109/95, 2P03B 244                                
          08p02, p03, p04 and p05, and the Foundation for Polish-German                            
          Collaboration (proj. No. 506/92) \\                                                      
$^{k}$ &  supported by the Polish State Committee for Scientific                                   
          Research (grant No. 2 P03B 083 08) and Foundation for                                    
          Polish-German Collaboration  \\                                                          
$^{l}$ &  partially supported by the German Federal Ministry for                                   
          Education and Science, Research and Technology (BMBF)  \\                                
$^{m}$ &  supported by the German Federal Ministry for Education and                               
          Science, Research and Technology (BMBF), and the Fund of                                 
          Fundamental Research of Russian Ministry of Science and                                  
          Education and by INTAS-Grant No. 93-63 \\                                                
$^{n}$ &  supported by the Spanish Ministry of Education                                           
          and Science through funds provided by CICYT \\                                           
$^{o}$ &  supported by the Particle Physics and                                                    
          Astronomy Research Council \\                                                            
$^{p}$ &  supported by the US Department of Energy \\                                              
$^{q}$ &  supported by the US National Science Foundation \\                                       
\end{tabular}                                                                                      
                                                           %
                                                           %

\newpage

\setlength{\dinwidth}{21.0cm} 
\textheight23.0cm 
\textwidth17.0cm
\setlength{\dinmargin}{\dinwidth}
\addtolength{\dinmargin}{-\textwidth}
\setlength{\dinmargin}{0.\dinmargin} 
\oddsidemargin -0.2in
\addtolength{\oddsidemargin}{\dinmargin}
\setlength{\evensidemargin}{\oddsidemargin}
\setlength{\marginparwidth}{0.9\dinmargin} 
\marginparsep 8pt
\marginparpush 5pt 
\topmargin -42pt 
\headheight 12pt 
\headsep 30pt
\footheight 12pt 
\footskip 24pt
\parskip 3mm plus 2mm minus 2mm
\parindent 0mm

\setcounter{page}{1}
\renewcommand{\thepage}{\arabic{page}}

\newpage
\section{Introduction} 
Jet photoproduction at HERA has been used to investigate various
aspects of quantum chromodynamics (QCD) and the structure of the
photon and the proton [1--6].
QCD calculations of jet cross sections can be factorized into two parts: 
the parton distributions in the beam particles and the matrix elements of 
the partonic hard scattering.
The selection of the kinematic region and variables in which dijet 
cross sections are studied can enhance the sensitivity of the data 
either to the matrix elements or to the parton distributions.

In a previous analysis~\cite{samerapdijet} dijet cross 
sections were measured in the regime where the difference between the 
pseudorapidities\footnote{$\eta = -$ln$(\tan\frac{\theta}{2}$) 
where $\theta$ is the polar angle with respect to the $z$ axis,
which in the ZEUS coordinate system is defined to be the proton direction.} 
of the two jets of highest transverse energy ($\ETJ$) is 
small ($|\eta^{jet1} - \eta^{jet2}| < 0.5$).
This constrained $\theta^{\ast}$, the angle between the 
jet-jet axis and the beam axis in the dijet centre of mass 
system, to be close to $90^o$. 
The cross section as a function 
of $\ETAB = (\eta^{jet1} + \eta^{jet2})/2$
was then sensitive to the parton distributions in the photon and proton.
In contrast, in the present analysis a cut is made on $\ETAB$,
resulting in a $\cts$ distribution which is sensitive to the 
parton dynamics.
The variable $\cts$ is calculated as
\begin{equation}
\cts = \tanh(\frac{\eta^{jet1} - \eta^{jet2}}{2}).
\end{equation}
Only the absolute value of $\cts$ can be determined
because the outgoing jets 
are indistinguishable. Measuring the distribution in $\cts$ is preferable to measuring 
the jet pseudorapidity distribution because $\cts$ is approximately
invariant under the different boosts along the beam axis arising 
from the spectrum of 
incoming parton momenta. This minimizes the sensitivity of the 
cross section to the momentum density distribution of the partons in 
the beam particles. 

In leading order (LO) QCD, two types of processes lead to the 
photoproduction of jets. 
In direct processes (Fig.1a) the photon participates in 
the hard scatter via either boson-gluon fusion or QCD Compton scattering. 
These processes involve a quark propagator in the 
$s$, $t$ or $u$ channel, with $t$ and $u$ channel processes dominating.
In resolved processes (Fig.1b) the photon acts as a source of 
quarks and gluons, and only a fraction of its momentum
participates in the hard scatter. In this case the dominant subprocesses, 
e.g. $q g \rightarrow q g$, $g g \rightarrow g g$ and 
$q q \rightarrow q q$,
have $t$-channel gluon exchange diagrams.
The angular dependence of the cross section for resolved 
processes with a spin-1 gluon propagator is approximately 
 $\propto ( 1 - |\cts| )^{-2}$ (as in Rutherford scattering). 
This cross section rises more steeply 
with increasing $|\cts|$ than that for direct processes with a 
spin-$\frac{1}{2}$ quark propagator, where the angular dependence 
is approximately 
$\propto  ( 1 - |\cts| )^{-1}$.
After inclusion of all LO diagrams, QCD predicts that the
angular distribution of the outgoing partons in resolved 
processes will be enhanced at high $|\cts|$ with respect to direct 
photon processes. 
This property is expected to be preserved in next-to-leading order
(NLO) calculations~\cite{owens}. 
In addition to depending upon the incoming flux of partons, this
prediction is sensitive to the relative colour factors for each
subprocess and to the spins of the quark and the gluon.

The separation between direct and resolved photoproduction
is only well defined at LO. To be able to make a measurement 
which can be compared to calculations at any order, the 
variable $\xgo$ is used to define these two kinematic 
regions~\cite{samerapdijet}.
The variable $\xgo$ is the fraction of the photon energy contributing
to the production of the two highest $\ETJ$ jets. It is defined as
\begin{equation}
  \xgo = \frac{E_{T}^{jet1}e^{-\eta^{jet1}} + E_{T}^{jet2}e^{-\eta^{jet2}}}
{2E_\gamma}
\end{equation}
where $E_\gamma$ is the initial photon energy. 
Direct processes as defined at leading order have high $\xgo$ since all the 
photon's energy participates in the production of the hard jets, while resolved 
processes as defined at leading order have low $\xgo$ since part of the photon's 
energy goes into
the photon remnant. 
Thus the different spins of the quark and gluon propagators that 
are dominant in the high $\xgo$ and the 
low $\xgo$ processes respectively 
should be reflected in the $\cts$ distributions of the two samples. 
Henceforth throughout the following, in both the data and the calculations, 
direct and resolved samples are defined in terms of a cut 
on $\xgo$ rather than in terms of the LO diagrams, unless explicitly stated 
otherwise.

Measurements of dijet angular distributions in $p\bar{p}$ events 
have shown good agreement with the predictions of perturbative 
QCD~\cite{ppcos} in both fermionic and bosonic exchange 
processes~\cite{ppcos2}. 
The measurement presented here tests QCD in a different
kinematic regime and in a different process, and provides an 
opportunity to study the parton dynamics of two distinct 
dijet production mechanisms in the same experiment.

\section{Experimental Setup}
In 1994 HERA provided 820~GeV protons and 27.5~GeV positrons 
colliding in 153 bunches. 
Additional unpaired positron and proton bunches circulated to 
allow monitoring of the background from beam-gas interactions. 
Events from empty bunches were used to estimate the background 
from cosmic rays. 
The total integrated luminosity used in this analysis is 
2.6~pb$^{-1}$ collected during this running period.

Details of the ZEUS detector have been described elsewhere~\cite{ZEUS}.  
The primary components used in this analysis are the central calorimeter
 and the central tracking detectors. 
The uranium-scintillator calorimeter~\cite{CAL} covers 99.7\% of the 
total solid angle and is subdivided into three parts, forward (FCAL) 
covering $4.3 > \eta > 1.1$,  barrel (BCAL) 
covering the central region $1.1 > \eta > -0.75$ and rear (RCAL) covering 
the backward region $-0.75 > \eta > -3.8$ for an event at the nominal 
interaction point.
Each part consists of an electromagnetic section followed by a hadronic 
section, with cell sizes of approximately 
$5 \times 20$ cm$^2$ ($10 \times 20$ cm$^2$ 
in the rear calorimeter) and $20 \times 20$ cm$^2$ respectively.
The central tracking system consists of a vertex detector~\cite{VXD} 
and a central tracking chamber~\cite{CTD}  enclosed in a 1.43~T solenoidal
 magnetic field. 
A lead-scintillator photon calorimeter is used to measure the luminosity 
via the positron proton Brems\-strahlung process. 
This calorimeter is installed 100 m along the HERA tunnel from the 
interaction point in the positron direction and subtends a small angle 
at the interaction vertex~\cite{LUMI}. 
Positrons scattered through small angles are detected in a similar lead-scintillator 
calorimeter. 

\section{Event Selection} 
The ZEUS detector uses a three level trigger system.
At the first level events were triggered on a coincidence of a 
regional or transverse energy sum in the calorimeter 
and a track from the interaction point measured in the central 
tracking chamber. 
At the second level at least 8~GeV total transverse energy, 
excluding the eight calorimeter towers immediately surrounding 
the forward beampipe, 
was required, and cuts on calorimeter energies and timing were used 
to suppress events caused by interactions between the proton beam 
and residual gas in the beam pipe~\cite{F2}. 
At the third level, jets were found using the calorimeter 
cell energies and positions as input to a cone algorithm\cite{snow}.
 Events were required to have at least two jets of $\ETJ > 3.5$~GeV 
and $\ETAJ < 3.0$. Additional tracking cuts were made to reject 
proton beam-gas interactions and cosmic ray events.

Further cuts are applied offline. Charged current events are rejected by a cut on
the missing transverse momentum measured in the calorimeter. To reject  
remaining beam-gas and cosmic ray background events, tighter cuts using the 
final vertex position, other tracking information and timing information 
are applied.
Two additional cuts are made~\cite{ZEUSdirect}, 
based upon different measurements of the inelasticity,
defined in the ZEUS frame as 
$y = 1 - \frac{\EEP}{2 E_e} (1-\cos{\TEP})$ where $E_e$ 
is the energy of the incoming positron and $\EEP$ and $\TEP$ are the 
energy and polar angle of the outgoing positron. 
\begin{enumerate}
\item 
Events with a positron candidate in the uranium calorimeter 
are removed if $y_e  < 0.7$, where $y_e$ is the value of $y$ 
as measured assuming the positron candidate is the scattered 
positron.
\item 
A cut is made on the Jacquet-Blondel measurement of $y$~\cite{YJB}, 
$y_{JB} = \sum_i (E_i - E_{zi}) /2E_e$, 
where $E_{zi} = E_i \cos\theta_i$, and 
$E_i$ is the energy deposited in the calorimeter cell $i$
which has a polar angle $\theta_i$ with respect to the 
measured $z$-vertex of the event.
The sum runs over all calorimeter cells.
For any event where the scattered positron entered the 
uranium calorimeter and either was not identified or gave 
$y_e$ above $0.7$, the value of $y_{JB}$ will be near to one. 
Proton beam-gas events will have low values of $y_{JB}$. 
To further reduce the contamination from both these sources,
it is required that $0.15 < y_{JB} < 0.7$.
This range corresponds to the true $y$ range of $0.25 < y < 0.8$.

\end{enumerate}
These cuts restrict the range of the photon virtuality 
to less than $\sim4$~GeV$^{2}$, with a median
of around $10^{-3}$~GeV$^2$, which excludes deep inelastic scattering 
(DIS) events.

To select dijet candidates, a cone algorithm is applied to the 
calorimeter cells using a cone  radius 
$R = \sqrt{(\delta\phi)^2 + (\delta\eta)^2} = 1$,
where $\delta\eta (\delta\phi)$ is the distance in $\eta (\phi)$ 
of the centre of a calorimeter cell from the jet centre.
The jet energy measured in the ZEUS detector has been corrected using 
the Monte Carlo(MC) events described in the next section. 
The energy response of the ZEUS calorimeter is estimated by 
comparing jets found 
in the hadronic final state of the MC generator to jets found in 
the simulated detector.
The average shift in jet energies is $-17\%$ and varies between
$-10\%$ and $-25\%$ depending 
upon $\eta^{jet}$.
The description of the jet energy shift in the MC has been checked using 
tracking and the incident photon
energy deduced from the energy of the electron which is 
measured in the small-angle electron tagger. This description 
is accurate to within 5\%~\cite{ZEUSinclusive,samerapdijet}. 
After this correction, 
events are required to have at least two jets with $\ETJ \ge 6$~GeV 
and $\ETAJ <2.5$. 
For events with three or more jets, the two highest $\ETJ$ jets are used
to calculate all jet-related event properties. This procedure is also 
employed later in all the theoretical and MC predictions shown.

For a given centre-of-mass energy, events at high $|\cts|$ have smaller scattering
 angles and thus lower $\ETJ$. 
In order to study the $|\cts|$ distribution up to $|\cts| = 0.85$ without 
bias from the $\ETJ$ requirement, a cut of $\MJJ > 23$~GeV has been 
applied. $\MJJ$ is the dijet invariant mass calculated using the relationship
\begin{equation}
\MJJ = \sqrt{ 2 E_T^{jet1} E_T^{jet2}[\cosh(\eta^{jet1} - \eta^{jet2}) - \cos(\phi^{jet1} -\phi^{jet2})]},
\end{equation}
where $\phi^{jet}$ is the azimuthal angle of the jet 
in the HERA frame. For two jets back-to-back in $\phi$ 
and with equal $\ETJ$, 
$\MJJ \approx 2 \ETJ / \sqrt{1 - |cos\theta^{\ast}|^2}$. 

Because of the asymmetric beam energies, the two parton 
centre-of-mass system is typically boosted to the forward direction.
For resolved photon processes, this effect is 
bigger
because only a 
fraction of the photon's momentum participates in the hard scatter.
MC simulations show that as a result, a significant fraction of jets 
are at rapidities ($\eta > 2.5$) where they are not well measured in 
the ZEUS detector.
To remove this bias, a cut $|\ETAB| < 0.5$ was applied, where $\ETAB$, the 
average $\eta$ of the two jets, is a measure of the boost of the dijet 
scattering system in the HERA frame: 
\begin{equation}
   \ETAB  \approx \eta_{boost} 
          = \frac{1}{2} ln(\frac{E_p x_p}{E_{\gamma} x_{\gamma}})
\end{equation}
where $x_p$ and $x_\gamma$ are the momentum fractions of the incoming 
partons in the proton and photon respectively and $E_p$ is the 
incoming proton energy.
In a simple (LO) $2 \rightarrow 2$ scatter, the dijet invariant 
mass is related to $x_p$ and $x_{\gamma}$ by
\begin{equation}
\MJJ = \sqrt{4 E_{\gamma} E_p x_{\gamma} x_p} = \sqrt{4 E_e E_p y x_{\gamma} x_p} .
\end{equation}
The requirement that the dijet system has high mass and small boost
selects events with $\gamma p$ centre-of-mass energies mostly 
above 190~GeV, and suppresses events with low $\xgo$.
Thus the incoming partons have approximately equal and opposite 
momenta and 
energies sufficient to produce dijets in a region of good acceptance 
for a wide range of scattering angles.
In the range of $\xg$ and $x_p$ selected by these cuts the 
photon and proton parton distributions are fairly well determined 
by previous 
measurements~\cite{H1inclusive,ZEUSinclusive,samerapdijet,F2,moreF2S}.

After these cuts a total of 4964 events remain with $|\cts| < 0.85$, of which
1982 have $\xgo < 0.75$ and 2982 have $\xgo \ge 0.75$.
Events with a third jet which passes the $\ETJ$ and $\eta^{jet}$ cuts 
comprise 9\% of the final sample.
21\% of all events have their scattered positrons detected in the
  small-angle tagger, as is expected for photoproduction events~\cite{ZEUSinclusive}. 
The beam-gas background, measured using unpaired bunches, 
is less than 1\%. 
The contamination from events with photon virtualities greater than 
4~GeV$^2$ is estimated  to be 3.3\% using simulated deep inelastic 
scattering (DIS) events. 

\section{Results and Discussion} 

In Fig.2 the ZEUS dijet data (black dots) are compared to 
the results of two QCD MC simulation programs, 
HERWIG58~\cite{HERWIG} (solid lines) and PYTHIA57~\cite{PYTHIA} 
(broken lines).
The GRV~\cite{GRV} parton distributions are used for the photon 
and the MRSA~\cite{MRSA} parton distributions are used for the proton.
The simulation programs are based on LO QCD calculations for the hard scatter
and include parton showering and hadronization effects. 
The minimum transverse momentum of the 
hard scatter is set to 2.5 GeV. For both programs the direct and 
resolved photon processes
are generated separately and combined according to the ratio of their
generated cross sections. 
The combined MC distributions are then normalized to the number of data events.
In the case of HERWIG resolved processes, multiparton interactions
are included~\cite{MI}, as this has been shown to improve 
the simulation of the energy flow around the jet axis in low-$\MJJ$ 
events~\cite{MIexp}.
All the MC events are passed through a detailed simulation of the 
ZEUS detector and the same jet energy correction procedure as was
applied to the data.
Fig.2a shows the distribution of events in $\ETJ$, which falls steeply with
increasing $\ETJ$. At low $\ETJ$ the dijet mass cut 
causes the distribution to turn over.
Fig.2b shows the $\MJJ$ distribution, which falls steeply with increasing
$\MJJ$ and extends to masses of 60~GeV.
The $\ETAB$ distribution is shown in Fig.2c, and 
rises with increasing $\ETAB$ due to the asymmetric momenta of the 
photon and proton.
The $\ETAB$ cut leads to an $\ETAJ$ distribution which is peaked in 
the forward and rear 
directions, as shown in Fig.2d. Note also that due to the
$\ETAB$ cut,  the absolute value of $\ETAJ$ is restricted 
to be below 1.8.
For all these distributions the data are reasonably well described 
by PYTHIA and HERWIG, although HERWIG gives more 
jets in the central region in $\ETAJ$ and at high $\ETJ$ than
are seen in the data.

Fig.3a shows the uncorrected $\xpo$ distribution, where $\xpo$
is the fraction of the proton's energy contributing to the production of the
two highest $\ETJ$ jets: 
\begin{equation}
\xpo = \frac{ E_{T1} e^{\eta^{jet1}} + E_{T2} e^{\eta^{jet2}}}{2E_p}.
\end{equation}
The events lie in the range $0.006 < \xpo < 0.06$, with the most probable
value at $\xpo \approx 0.016$. 
Fig.3b shows the uncorrected $\xgo$ distribution\footnote{$\xgo$ was 
measured using the uncorrected jet transverse energies. This takes 
advantage of cancellations between energy losses in 
$\ETJ$ and $E_\gamma = y_{JB}E_e$.}.
Contributions from the direct photon LO diagrams (e.g. Fig.1a)  
(broken line) and the resolved photon LO diagrams (e.g. Fig.1b) 
(dotted line) of HERWIG are shown together with the combined 
distribution (solid line).
PYTHIA events, not shown in the figure, have a similar distribution.
The data exhibit a slightly lower mean $\xgo$ than the MC.
A cut is applied at $\xgo = 0.75$ to define the direct and resolved 
photon samples. The effect of the imperfect description of the 
data by MC on the evaluation of the cross sections is evaluated by
varying the $\xgo$ cut, and is small.
The transverse energy flow around the jets is shown in 
Fig.3c for resolved and in Fig.3d for direct events.
In this sample, which has high $\MJJ$ values, 
both the HERWIG sample with
multiple interactions included and the PYTHIA sample without 
multiple interactions describe the jet profiles 
reasonably well in both direct and resolved events.
The requirements of high mass and small boost remove the 
disagreement in the forward energy flow between data and the simulations 
which has been reported elsewhere~\cite{H1inclusive,ZEUSinclusive,samerapdijet}
in hard photoproduction at HERA. 

The resolution of the kinematic variables has been studied
by comparing in the MC simulation jets reconstructed from 
final state particles 
(hadron jets) with jets reconstructed from the energies measured in the 
calorimeter (calorimeter jets), and by comparing $y_{JB}$ with the true $y$.
The difference in $|\cts|$ between the hadron 
and detector levels has a mean of zero, a width of 0.03 and is approximately 
independent of $|\cts|$, $\MJJ$ and $\ETAB$. 
The resolution of $\xgo$ is 8.7\% at $\xgo = 0.75$. 
For the variables $\ETJ$, $\MJJ$ and $y$, the
resolutions are 13.2\%, 11.1\% and 11.0\% 
respectively at the values at which the cuts are applied. 

The MC samples have been used to correct for the efficiency of 
the trigger and selection cuts and migrations, and for the contamination from 
DIS events.
The final bin-by-bin correction factors are approximately
independent of $|\cts|$ and are around 1.1 for the resolved 
and 1.5 for the direct cross section. These correction factors 
are calculated as the ratio of the purity ($=N_{\rm true, rec}/N_{\rm rec} $)
to the efficiency ($= N_{\rm true,rec}/N_{\rm true}$) in each bin.
$N_{\rm true}$ is the number of events generated in the bin,
$N_{\rm rec} $ is the number of events reconstructed in the bin after 
detector smearing and all experimental cuts, and 
$N_{\rm true, rec}$ is the number of events which were both 
generated in the bin and reconstructed in that bin.
The difference between the correction factors for resolved and direct
events is due to the fact that the $\ETJ$ and $\MJJ$ distributions are 
steeper in resolved processes than in the direct, and thus migrations 
from below the cuts are more significant, giving a lower purity for
resolved than direct.

The sensitivity of the measured cross sections to the selection cuts 
has been investigated by varying the cuts on the reconstructed variables
in the data and HERWIG MC samples and re-evaluating the cross section.
In addition, the cross section was re-evaluated using a different ratio
of the direct and resolved contributions, and using the PYTHIA sample.
The systematic uncertainty arising from the subtraction of the
contamination coming from DIS was estimated by using two different DIS Monte Carlo 
generators and two different positron-finding algorithms~\cite{F2,moreF2S}.
We have also allowed for the possibility that the detector
simulation may overestimate the
detector energy response by up to 5\%, as mentioned in section~3.

The cross sections $d\sigma/d|\cts|$ for 
$ep \rightarrow$ dijets $+ X$ in the range
$\MJJ > 23$~GeV, $\ETJ > 6$~GeV, $|\ETAB| < 0.5$,
$0.25 < y < 0.8$ and for virtualities of the
exchanged photon $< 4$~GeV$^2$ 
are listed in table~1. 
For the direct photoproduction sample the 
$ep$ cross section rises from around 0.8~nb
at $|\cts |= 0$ to around 6~nb at $|\cts| = 0.85$. 
For the resolved photoproduction sample 
the $ep$ cross section rises from around 
0.2~nb at $|\cts| = 0$ to around 4~nb at $|\cts| = 0.85$. 
In Fig.4a and b the two cross sections have been
normalized to unity at $|\cts| =0$ so that the shapes can be compared directly.
To determine the normalization, the 5 lowest $|\cts|$ bins were fitted
to the functional form expected for the dominant cross section in each sample
(as described in section~1), which is slowly rising
in this region. The only free parameter in this fit is the normalisation of the
function. The intercept of the function then gives the factor by which the
data are divided. 
In Fig.4a the data are compared to LO and partial~\footnote{The NLO calculations 
do not yet include terms of {\cal O}($\alpha_s^3 \times f$), where $f$
is the flux of partons in the photon.} 
NLO QCD parton level calculations~\cite{owens}. 
The resolved cross section is seen to rise more steeply than
the direct cross section with increasing $|\cts|$.
There is good agreement between data and theory, verifying the
expected effects of the spins of the quark and gluon propagators.
This is also seen in Fig.4b, where the data are compared
to the HERWIG and PYTHIA predictions. For both of the generators,
MRSA(proton) and GRVLO(photon)~\cite{GRV} parton parameterisation 
sets are used.
The MC curves were also made with different parton 
distribution sets in 
the photon and proton. The largest 
change seen was for the LAC1~\cite{LAC1} 
photon parton distribution set, where there is a 35\% increase 
in the prediction in the highest $|\cts|$ bin.
In all the calculations, resolved and direct samples are defined on the
basis of the cut on $\xgo$, exactly as in the data.
Both the MC and the parton level calculations show the same behaviour, 
suggesting that the effects of parton showers and hadronisation on this 
distribution are small. The data are in good agreement 
with the  MC and with the calculations.

The only difference in the cuts applied to the two samples is the 
cut on $\xgo$ which is used to separate the resolved and direct samples.
A MC study has been performed to investigate whether the different behaviour 
of the two cross sections might be a kinematic effect of this cut.
The study shows that the tail of the resolved photon LO diagrams 
(e.g. Fig.1b) which lie above $\xgo = 0.75$ retain the
steeply rising $|\cts|$ distribution characteristic of the low-$\xgo$ events.
Thus the different behaviour of the two cross sections is due to the different partonic 
subprocesses which contribute in the different $\xgo$ regions.

\section{Summary and Conclusion} 
The dijet angular cross section $d\sigma/d|\cts|$, where $\theta^*$ is
the jet scattering angle in the dijet c.m.s,  
has been measured for $ep \rightarrow 2$ or more jets 
with transverse energies $E_T^{jet} \ge$ 6 GeV, dijet invariant mass 
$\MJJ > 23$~GeV and  average pseudorapidity $|\ETAB| < 0.5$, 
in the range $0.25 < y < 0.8$ and for virtualities of the exchanged photon
less than 4 GeV$^2$.
The cross section has been measured for resolved ($\xgo < 0.75$) and
direct ($\xgo \ge 0.75$) processes. The angular dependence 
for the two samples is significantly different.

The dependence of $d\sigma/d|\cts|$ on $|\cts|$ reflects the 
different spins of the quark and gluon propagators and the 
relative contributions
of the underlying subprocesses in resolved and direct photoproduction. 
In LO QCD the cross section rises faster with increasing 
$|\cts|$ for resolved photoproduction, where processes involving
spin-1 gluon exchange dominate, than for direct photoproduction, 
where processes
involving spin-1/2 quark exchange dominate.
These expectations are preserved in NLO QCD calculations and in Monte Carlo 
simulations which include parton showering and hadronisation models.
The $|\cts|$ dependence of the measured cross sections is in good agreement
with these theoretical predictions and thus confirms
fundamental aspects of quantum chromodynamics.

\section*{Acknowledgements}
It is a pleasure to thank the HERA accelerator group and the computing 
and networking support staff at DESY, without whom this analysis 
would not have been possible, and to thank J. Owens for providing us 
with his calculations. The continuing encouragement from
the DESY directorate is greatly appreciated.

\clearpage
\newpage

\begin{table}[t]
\centering
\begin{tabular}{|c| c|c|c|c|c|}  \hline
$|\cts|$      & \multicolumn{5}{c|}{ Direct (nb / unit $|\cts|$) } \\ \cline{2-6} 
range                   & $d\sigma/d|\cts| \pm$  stat. & syst. error & PYTHIA & E scale --5\% & E scale +5\% \\ \hline
0.000 to 0.085          &  0.80  $\pm$0.12 & +0.06/--0.09 & -0.08 & -0.07 & 0.06 \\
0.085 to 0.170          &  0.83  $\pm$0.13 & +0.07/--0.13 & -0.10 & -0.13 & 0.11 \\
0.170 to 0.255          &  0.98  $\pm$0.15 & +0.02/--0.13 & -0.17 & -0.16 & 0.16 \\
0.255 to 0.340          &  0.90  $\pm$0.14 & +0.02/--0.08 & -0.13 & -0.12 & 0.12 \\
0.340 to 0.425          &  1.26  $\pm$0.19 & +0.15/--0.16 & -0.30 & -0.21 & 0.26 \\
0.425 to 0.510          &  1.50  $\pm$0.18 & +0.08/--0.19 & -0.26 & -0.26 & 0.24 \\
0.510 to 0.595          &  1.41  $\pm$0.17 & +0.18/--0.08 & -0.10 & -0.21 & 0.14 \\
0.595 to 0.680          &  2.70  $\pm$0.28 & +0.36/--0.35 & -0.27 & -0.53 & 0.51 \\
0.680 to 0.765          &  3.81  $\pm$0.33 & +0.14/--0.29 & -0.30 & -0.57 & 0.36 \\
0.765 to 0.850          &  5.99  $\pm$0.48 & +0.81/--0.40 & -0.37 & -0.96 & 0.86 \\ \hline \hline
$|\cts|$      & \multicolumn{5}{c|}{ Resolved (nb / unit $|\cts|$) } \\ \cline{2-6} 
range                   & $d\sigma/d|\cts| \pm$  stat. & syst. error & PYTHIA & E scale --5\% & E scale +5\% \\ \hline
0.000 to 0.085          & 0.21   $\pm$0.08 & +0.06/--0.08 & -0.09 & -0.04 & 0.00 \\
0.085 to 0.170          & 0.23   $\pm$0.12 & +0.09/--0.12 & -0.08 & -0.06 &-0.02 \\
0.170 to 0.255          & 0.23   $\pm$0.10 & +0.04/--0.07 &  0.01 & -0.06 & 0.12 \\
0.255 to 0.340          & 0.28   $\pm$0.09 & +0.13/--0.02 &  0.00 & -0.02 & 0.04 \\
0.340 to 0.425          & 0.27   $\pm$0.10 & +0.10/--0.06 & -0.03 &  0.01 & 0.02 \\
0.425 to 0.510          & 0.42   $\pm$0.11 & +0.11/--0.04 & -0.03 & -0.06 & 0.11 \\
0.510 to 0.595          & 0.77   $\pm$0.17 & +0.08/--0.08 &  0.21 & -0.13 & 0.10 \\
0.595 to 0.680          & 1.13   $\pm$0.21 & +0.16/--0.26 & -0.02 & -0.22 & 0.07 \\
0.680 to 0.765          & 2.03   $\pm$0.25 & +0.36/--0.19 &  0.01 & -0.39 & 0.63 \\
0.765 to 0.850          & 4.31   $\pm$0.43 & +0.47/--0.39 &  0.15 & -0.85 & 0.97 \\ \hline
\end{tabular}
\caption{Differential cross sections $d\sigma/d|\cts|$ for $ep \rightarrow 2$ or more jets 
with $\ETJ \ge 6$~GeV, $\MJJ > 23$~GeV and $|\ETAB| < 0.5$, 
in the range $0.25 < y < 0.8$ and for virtualities of the exchanged photon
less than 4 GeV$^2$, for direct and resolved processes.
The first column shows the cross section and statistical errors in units of nb,
the second column the uncorrelated systematic errors, the third column the
shift in the cross section when the correction is evaluated using PYTHIA instead of 
HERWIG and the fourth and fifth columns the shifts in the cross section when the calorimeter 
energy calibration is varied by $\pm 5\%$. There is an additional overall uncertainty 
of 3.3\% arising from the luminosity measurement.}
\end{table}

\clearpage
\newpage

\begin{figure}[t]
\vspace{-4cm}
\hspace{-1cm}
\setlength{\unitlength}{1mm}
\epsfysize=200pt
\epsfbox[200 400 500 800]{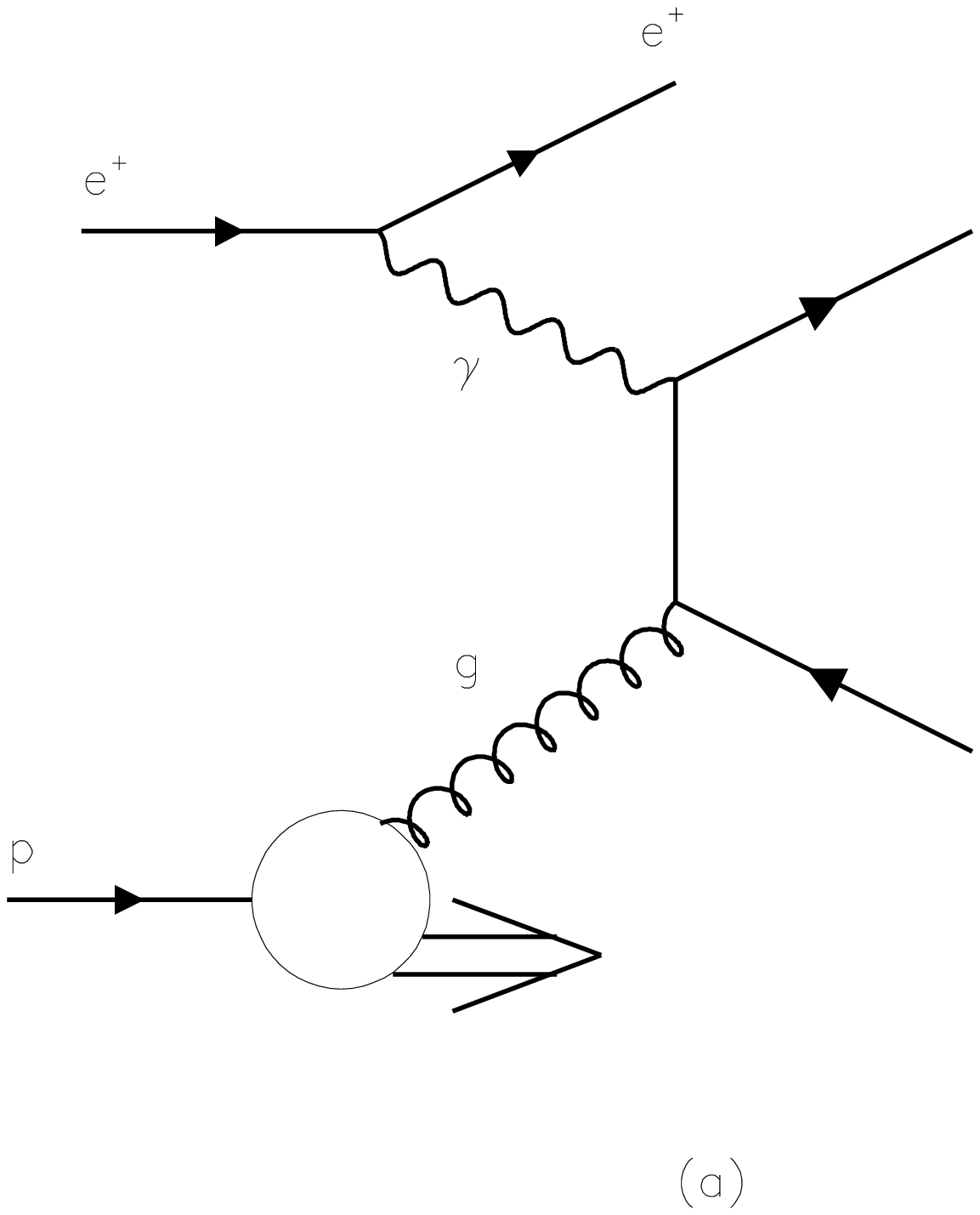}
\epsfysize=200pt
\epsfbox[0 400 300 800]{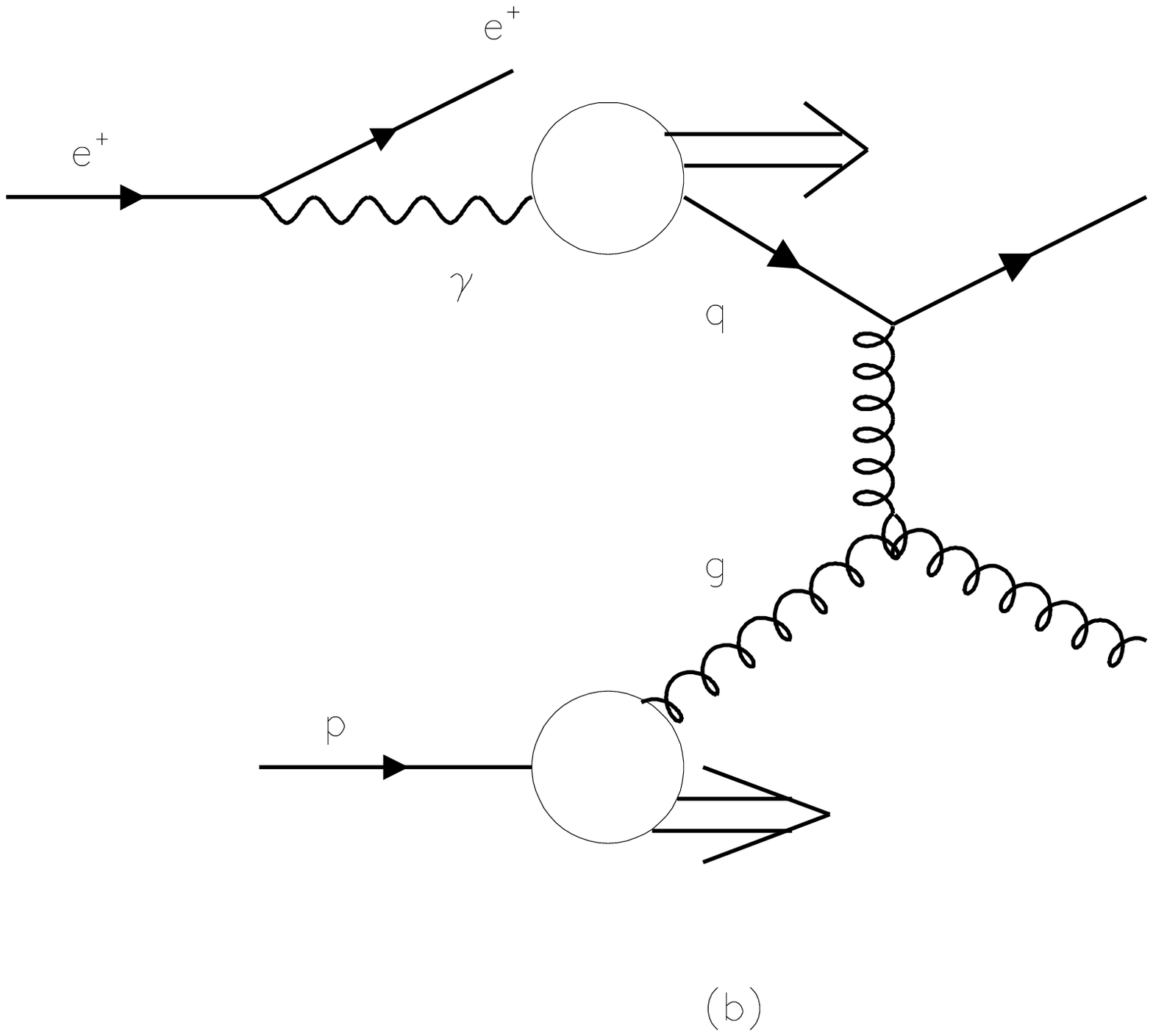}
\vspace{3.5cm}
\caption{Examples of LO QCD (a) `direct' and 
(b) `resolved' dijet production diagrams.}
\end{figure}

\clearpage
\newpage
\begin{figure}[t]
\vspace{-3cm}
\epsfxsize=14cm
\hspace{2cm}
\epsffile{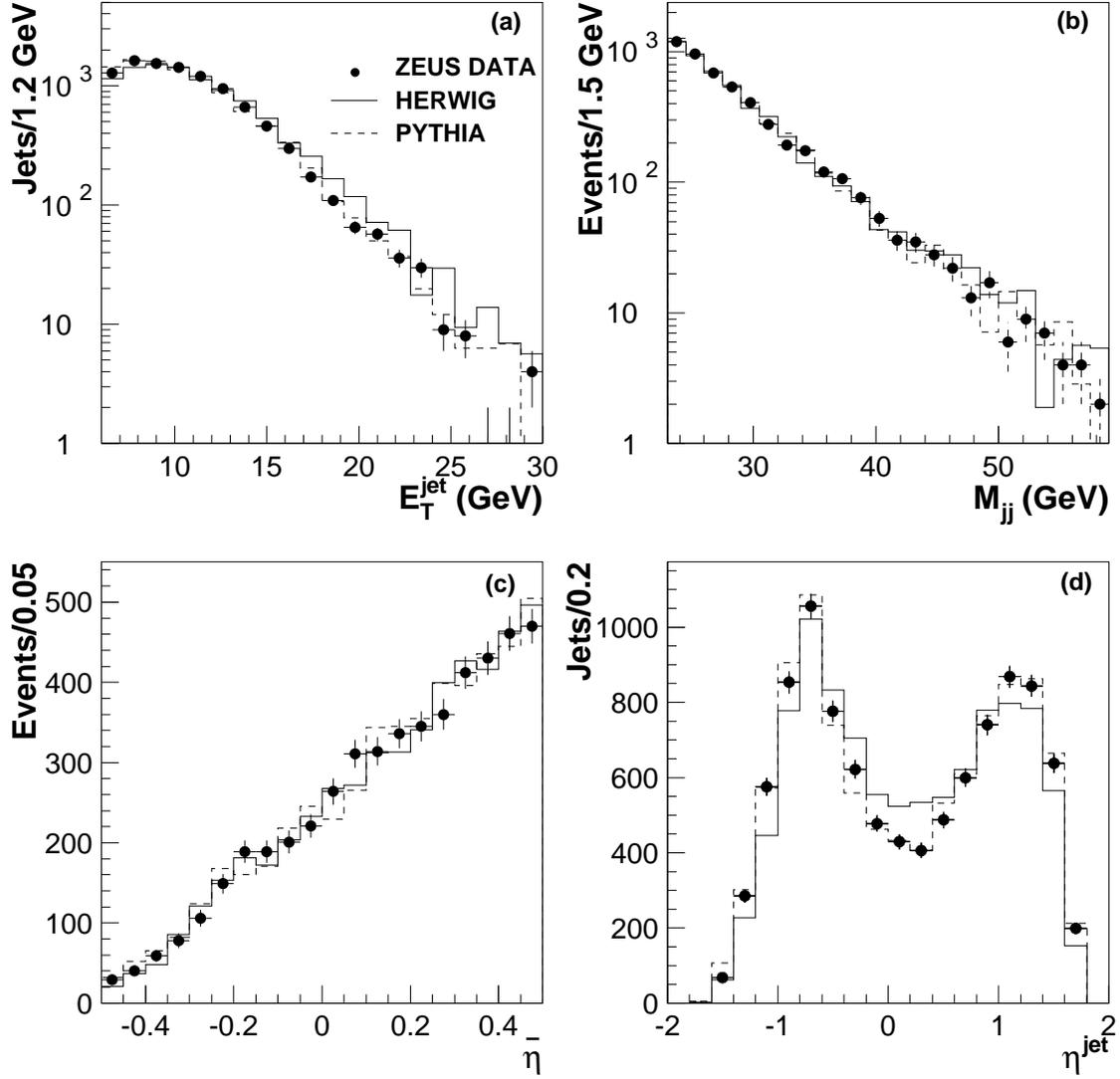}
\vspace{-1cm}
\caption{ (a)  $\ETJ$,  
          (b)  $\MJJ$, 
          (c)  $\ETAB$ and 
          (d)  $\ETAJ$
distributions. 
Raw ZEUS data are compared to the results of the 
HERWIG58 (solid line) and PYTHIA57 (broken line) Monte Carlo 
models after simulation of all detector effects and application of selection cuts. 
The MC samples have been normalized to the number of events in the
data.
Only statistical errors are shown.} 
\end{figure}

\clearpage
\newpage
\begin{figure}[t]
\vspace{-3cm}
\epsfxsize=14cm
\hspace{2cm}
\epsffile{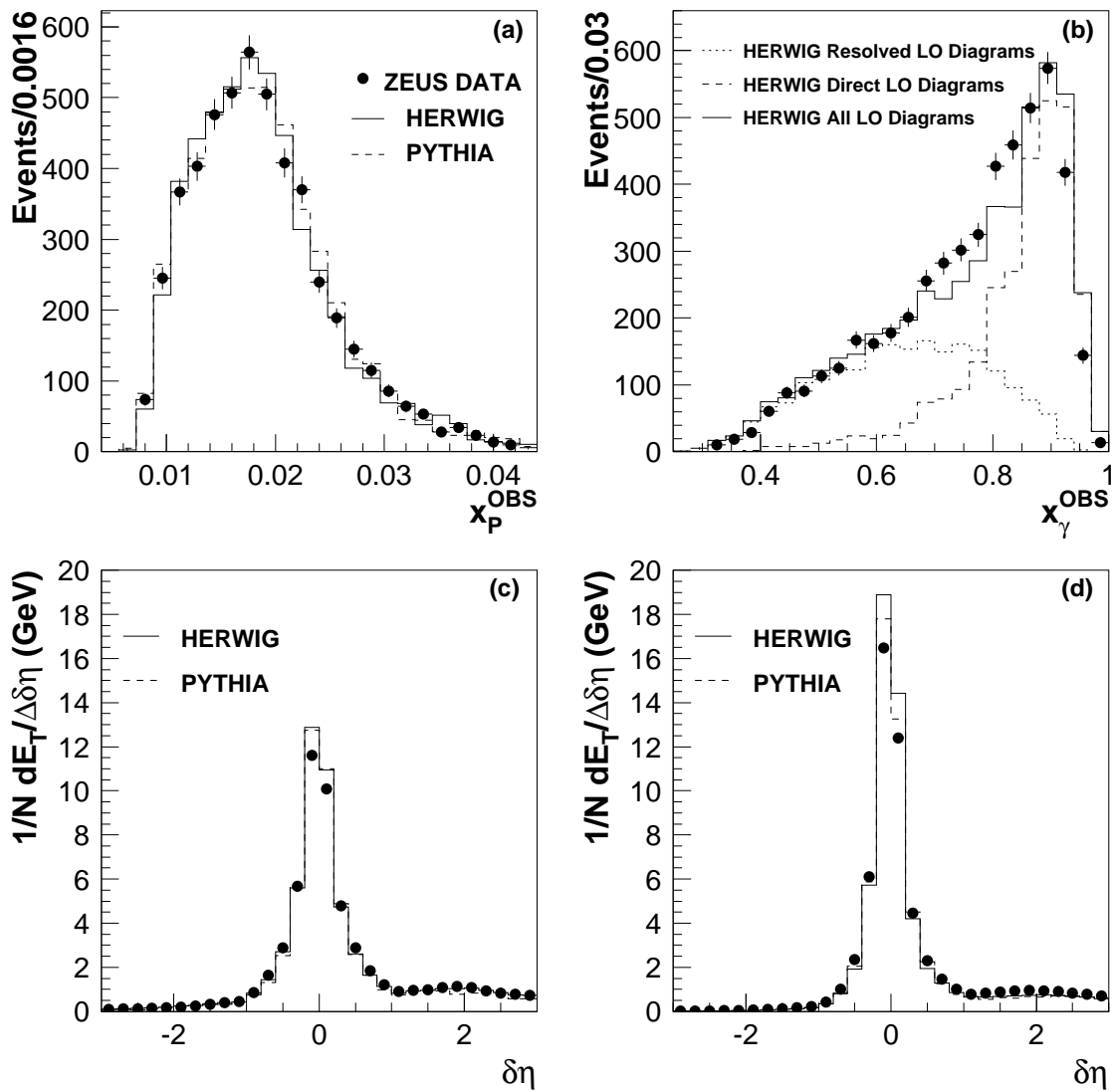}
\vspace{-1cm}
\caption{Raw ZEUS data compared to results of Monte Carlo models 
after simulation of detector effects and application of selection cuts.
 (a) $\xpo$ and  (b) $\xgo$ distribution. 
The MC samples have been normalized to the number of events in the
data.
Figures (c) and (d) show the uncorrected transverse energy 
flow $1/N dE_T/d\delta\eta$ around the jet axis, for cells within one 
radian in $\phi$ of the jet axis, for (c) resolved ($\xgo < 0.75$), 
(d) direct ($\xgo > 0.75$) events. 
Only statistical errors are shown.
In (a), (c) and (d), the histograms are HERWIG58 (solid line) and 
PYTHIA57 (broken line) 
In (b), the solid line is HERWIG58, the broken line is the distribution of
the LO direct diagrams (e.g. Fig.1a) and the dotted line is that of the
LO resolved diagrams (e.g. Fig.1b).}
\end{figure}

\clearpage
\newpage
\begin{figure}[t]
\vspace{-3cm}
\epsfxsize=14cm
\hspace{2cm}
\epsffile{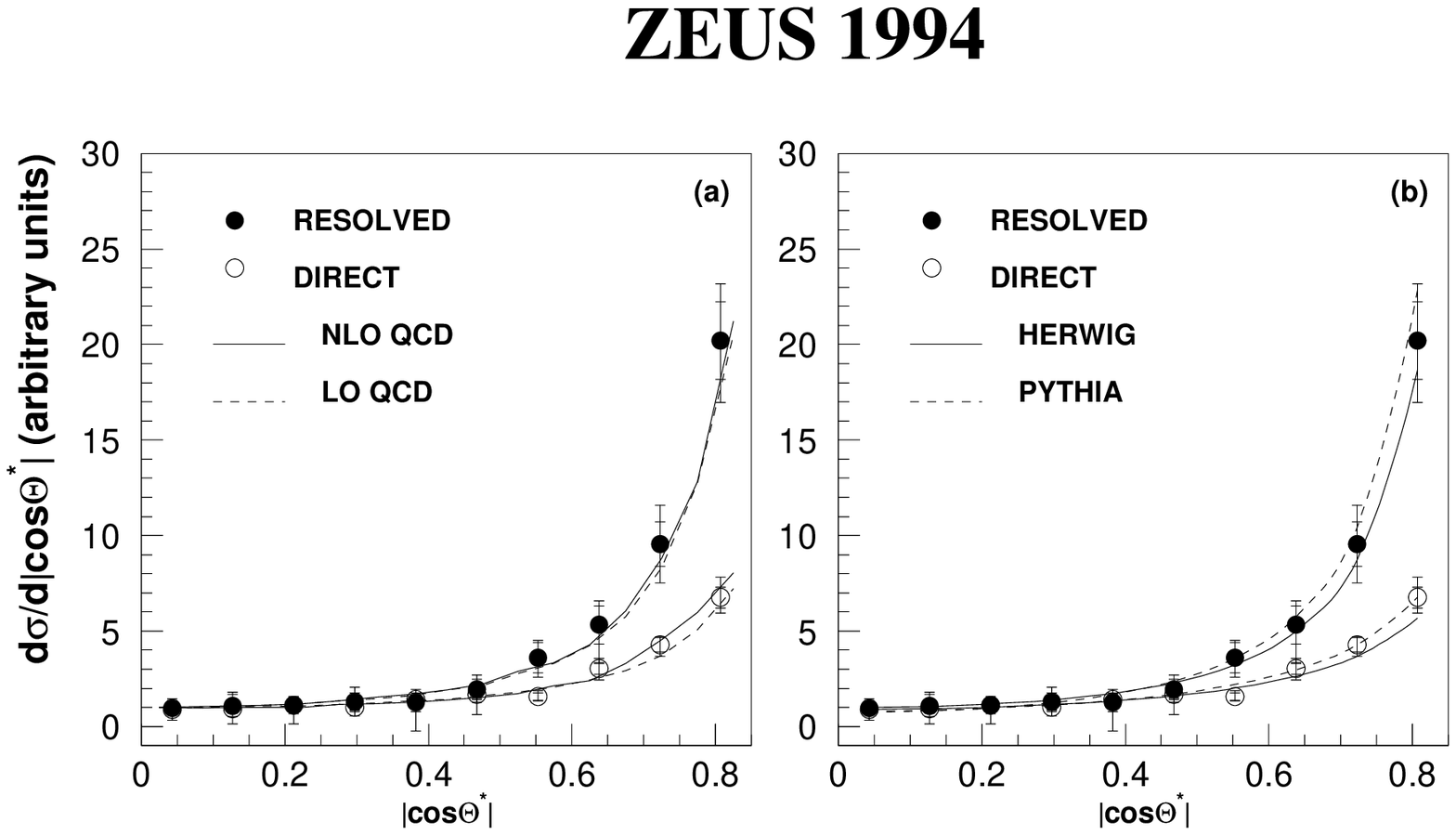}
\vspace{-9cm}
\caption{$d\sigma/d|\cts|$ normalized to one at $\cts = 0$
for resolved (black dots) and direct (open circles) photoproduction.
In (a), the ZEUS data are compared to the NLO prediction (solid line)  
and the LO prediction (broken line). The parton distribution sets
used in the calculation are CTEQ3M~[26] for the proton and GRV 
(LO)~[21] for the photon.
In (b), the broken line is the PYTHIA distribution and
solid line is HERWIG distribution. The inner error bars are the 
statistical errors,
the outer error bars are the sum in quadrature of the statistical and 
systematic errors,
excluding the energy scale and luminosity uncertainties (see Table~1).}
\end{figure}

\end{document}